\documentclass[12pt, onecolumn, draftclsnofoot]{IEEEtran}

\usepackage[noadjust]{cite}
\usepackage{amssymb, amsmath}
\usepackage{graphicx}

\newtheorem{theorem}{Theorem}
\newtheorem{lemma}[theorem]{Lemma}
\newtheorem{corollary}[theorem]{Corollary}
\newtheorem{definition}[theorem]{Definition}
\newenvironment{remark}{\par {\itshape Remark:}}{}

\newcommand{\beqn}{\begin{equation}}
\newcommand{\eeqn}{\end{equation}}
\newcommand{\beq}{\begin{equation*}}
\newcommand{\eeq}{\end{equation*}}
\newcommand{\Z}{\mathbb Z}
\newcommand{\Rn}{\mathbb R}

\newcommand{\B}{{\cal B}}
\newcommand{\C}{{\cal C}}
\newcommand{\D}{{\cal D}}

\newcommand{\RM}{{\rm RM}}
\newcommand{\PMEPR}{\mbox{PMEPR}}

\newcommand{\wt}{{\rm wt}}

\renewcommand{\QED}{\QEDopen}

\begin{document}
\date{}

\title{On the Peak-to-Mean Envelope Power Ratio of\\ Phase-Shifted Binary Codes}
\author{Kai-Uwe Schmidt\thanks{Kai-Uwe Schmidt is with Communications Laboratory, Dresden University of Technology, 01062 Dresden, Germany, e-mail: schmidtk@ifn.et.tu-dresden.de, web: http://www.ifn.et.tu-dresden.de/$\sim$schmidtk/}}%

\maketitle


\begin{abstract}
The peak-to-mean envelope power ratio (PMEPR) of a code employed in orthogonal frequency-division multiplexing (OFDM) systems can be reduced by permuting its coordinates and by rotating each coordinate by a fixed phase shift. Motivated by some previous designs of phase shifts using suboptimal methods, the following question is considered in this paper. For a given binary code, how much PMEPR reduction can be achieved when the phase shifts are taken from a $2^h$-ary phase-shift keying ($2^h$-PSK) constellation? A lower bound on the achievable PMEPR is established, which is related to the covering radius of the binary code. Generally speaking, the achievable region of the PMEPR shrinks as the covering radius of the binary code decreases. The bound is then applied to some well understood codes, including nonredundant BPSK signaling, BCH codes and their duals, Reed--Muller codes, and convolutional codes. It is demonstrated that most (presumably not optimal) phase-shift designs from the literature attain or approach our bound.
\end{abstract}


\section{Introduction}

\PARstart{O}{rthogonal} frequency-division multiplexing (OFDM), a special kind of multicarrier communications, is a key concept in the development of wired and wireless communications systems in the past decade. It provides excellent ability to cope with multipath propagation and fast-moving environment. On the other hand, a principal drawback of OFDM is the typically high peak-to-mean envelope power ratio (PMEPR) of uncoded OFDM signals. That is, the peak transmit power can be many times the average transmit power.
\par
In order to ensure a distortionless transmission, all components in the transmission chain must be linear across a wide range of signal levels. This makes the transmitter considerably more expensive than one in a single-carrier system. Moreover, most of the time, the components in the transmitter are operated at levels much below their maximum input level, which results in power inefficiency. The latter issue is particularly acute in mobile applications, where battery lifetime is a major concern. On the other hand, nonlinearities in  the transmission chain may lead to a loss of orthogonality among the carriers and to out-of-band radiation. The former has the effect of degrading the total system performance and the latter is subject to strong regulations.
\par
Among various approaches to solve this power-control problem, the use of block coding across the carriers \cite{Jones1994} combined with error protection \cite{Jones1996} is one of the most promising concepts \cite{Paterson2000}. Here the goal is to design error-correcting codes that contain only codewords with low PMEPR.
\par
A simple approach for the design of such codes was originally proposed by Jones and Wilkinson \cite{Jones1996} and further developed by Tarokh and Jafarkhani \cite{Tarokh2000}. The idea is to take a well understood code and to rotate each coordinate of the code by a fixed phase shift such that the maximum PMEPR taken over all codewords is minimized. This modification leaves unchanged the rate and the error-correcting properties of the code. Moreover a standard decoder for the original code can be employed in the receiver upon back rotation of the phase shifts. 
\par
Unfortunately, except perhaps for very short codes, the computation of the optimal phase shifts (which minimize the PMEPR) is considered to be extremely difficult and a feasible solution is unknown. Several suboptimal algorithms have been proposed in the literature \cite{Jones1996}, \cite{Tarokh2000}, \cite{Wunder2002}. These techniques were applied to obtain phase shifts for short Hamming codes \cite{Jones1996}, convolutional codes \cite{Tarokh2000}, \cite{Wunder2002}, and nonredundant signaling \cite{Wunder2002}. However the achieved PMEPR reductions are rather small. One reason for this may be the suboptimality of the employed algorithms: the use of suboptimal techniques generally results in the convergence of the algorithm to a local minimum, and therefore, it is never certain that the optimal phase shifts are computed.
\par
In this paper we study the fundamental limit of the achievable PMEPR of phase-shifted binary codes. We prove a lower bound for this limit and try to answer the question whether suboptimal phase-shift-design algorithms reported in the literature are in principle useful to approach this limit. The bound can be used to estimate the gap between the reduced PMEPR, obtained using suboptimal methods, and the global minimum. This allows us, in many cases, to establish the (near) optimality of some known phase-shift designs. We will also see that for several codes of practical importance a significant PMEPR reduction is ruled out by our results.
\par
An outline of the remainder of this paper is given below. In the next section we introduce a simple OFDM model and state the problem formally. This involves a quantization of the phase shifts such that the phase-shifted code is a $2^h$-ary phase-shift keying ($2^h$-PSK) code. Theorem~\ref{thm:LB_general} in Section~\ref{sec:main_result} states a lower bound on the PMEPR of such codes. The general case, where the restriction on the phase shifts is dropped, is then recovered in Corollary~\ref{cor:LB_arbitrary} by analyzing this bound for $h\to\infty$. The implication of this result for several codes, including nonredundant BPSK signaling, BCH codes and their duals, Reed--Muller codes, and convolutional codes, is discussed in Section~\ref{sec:implications}. Section~\ref{sec:proofs} contains the proofs of our main results. In Section~\ref{sec:conclusions} we close with some concluding remarks and a brief discussion of open problems.

\section{Problem Statement}
\label{sec:definitions}
 
Consider a code $\C\subseteq\Z_{2^h}^n$, and let $c=(c_0,\dots,c_{n-1})$ be a codeword in $\C$. The element $c_i$ is referred to as the $i$th \emph{coordinate} of $\C$. 
With any $c$ we associate another word $\hat c=(\hat c_0,\dots,\hat c_{n-1})$, where $\hat c_i=e^{j2\pi c_i/2^h}$ and  $j$ is an imaginary unit such that $j^2=-1$. Then $\{\hat c\,|\,c\in\C\}$ is a $2^h$-PSK code. Given a codeword $c\in\C$, the transmitted OFDM signal is the real part of the \emph{complex envelope}, which can be written as
\beqn
\label{eqn:OFDM_signal}
S_c(\theta)=\sum_{i=0}^{n-1}\hat c_i\,e^{j2\pi (i+\zeta)\theta},\qquad 0\le\theta<1,
\eeqn
where $\zeta$ is a positive constant. Note that the guard interval is omitted in our model since it does not affect the PMEPR of the signal. Moreover our model only considers codes defined over PSK constellations.
\par
The signal $|S_c(\theta)|^2$, which is independent of $\zeta$, is called the \emph{instantaneous envelope power} of $S_c(\theta)$. It is a consequence of Parseval's identity that the mean value of $|S_c(\theta)|^2$ is equal to $||\hat c||^2=\sum_{i=0}^{n-1}|\hat c_i|^2=n$. Therefore the PMEPR of the word $c$ (or of the signal $S_c(\theta)$) is given by
\beqn
\label{eqn:PMEPR_codeword}
\PMEPR(c):=\frac{1}{n}\sup_{0\le \theta<1}|S_c(\theta)|^2.
\eeqn
Occasionally we shall indicate the PMEPR of $c$ as $10\cdot\log_{10}\PMEPR(c)$~[dB]. The PMEPR of the code $\C$ is now defined to be
\beq
\PMEPR(\C):=\max_{c\in\C}\PMEPR(c).
\eeq
Observe that $\PMEPR(\C)$ can be as much as $n$, which occurs, for example, if $\C$ contains a constant codeword. Hence the PMEPR of every linear code is equal to $n$.
\par
The OFDM coding problem may now be stated as follows. Design codes of length $n$ with high rate, large minimum distance, and PMEPR significantly lower than $n$. This problem is considered to be difficult, and only a few explicit constructions are known \cite{Paterson2000}, \cite{Davis1999}, \cite{Paterson2000a}, \cite{Schmidt2006c}, \cite{Schmidt2007}. An even more challenging problem is the construction of sequences of codes with increasing length $n$, nonvanishing rate, and PMEPR growing strictly slower than linearly in $n$. Some results on the existence of such codes have been established in \cite{Paterson2000}, \cite{Sharif2004}.
\par
The following approach for designing codes with reduced PMEPR was originally proposed by Jones and Wilkinson \cite{Jones1996}.
\par
Let $\B\subseteq\Z_2^n$ be a binary code. We say that a code $\C\subseteq\Z_{2^h}^n$ is \emph{equivalent} to $\B$ if there exists a permutation $\sigma$ of $\{0,1,\dots,n-1\}$ and an offset $w\in\Z_{2^h}^n$ such that
\beq
\C=\left\{2^{h-1}\sigma(b)+w\,|\,b\in\B\right\},
\eeq
where we write $\sigma(b)$ in place of $(b_{\sigma(0)},\dots,b_{\sigma(n-1)})$. Note that, if $h=1$ and $\B$ is linear, $\C$ is permutation equivalent to a coset of $\B$. 
\par
The $2^h$-PSK code corresponding to $\C$ is given by
\beq
\left\{\hat b_{\sigma(0)} e^{j2\pi w_0/2^h},\dots,\hat b_{\sigma({n-1})} e^{j2\pi w_{n-1}/2^h}\,\big|\,b\in\B\right\}.
\eeq
Such a code is a \emph{phase-shifted version} of a code that is permutation equivalent to the BPSK code associated with $\B$, where the $i$th phase shift is equal to $2\pi w_i/2^h$ with $w_i\in\Z_{2^h}$. Notice that, when $h$ tends to infinity, any phase shift can be approximated in this way with arbitrarily high precision.
\par
Now let $E_h(\B)$ be the set of all codes $\C\subseteq\Z_{2^h}^n$ that are equivalent to $\B$. By $E_\infty(\B)$ we denote the infinite set $E_h(\B)$ when $h$ tends to infinity (that is, the codes in $E_\infty(\B)$ are defined over the $2$-adic integers $\Z_{2^\infty}$). Any code in $E_h(\B)$ has the same error-correcting capability and the same rate as $\B$, but the PMEPR of a particular $\C\in E_h(\B)$ may be much lower than that of $\B$.
\par
The power-control problem for OFDM can now be tackled as follows. Given a positive integer $h$ and a good (in the classical coding-theoretic sense) binary error-correcting code $\B$, find a code in $E_h(\B)$ whose PMEPR is equal to
\beqn
\label{eqn:min_PMEPR}
\min_{\C\in E_h(\B)}\PMEPR(\C).
\eeqn
This problem is considered to be extremely difficult, and only suboptimal algorithms are known, which may not even find a code whose PMEPR is close to the preceding expression. In this light we ask: what is the value of (\ref{eqn:min_PMEPR})? A lower bound for this quantity will be stated in the next section.
\par
We remark that, although (\ref{eqn:min_PMEPR}) is a nonincreasing function of $h$, in practice, $h$ is typically small, say $2$ or $3$, in order to allow efficient hardware implementation.


\section{Main Result}
\label{sec:main_result}

Recall that the \emph{Hamming weight}, $\wt_H(b)$, of a binary vector $b$ is equal to the number of nonzero elements in $b$. Our lower bound for (\ref{eqn:min_PMEPR}) will be expressed in terms of the covering radius of the binary code $\B$, which is defined below.
\begin{definition}
\label{def:covering_radius}
The \emph{covering radius} of a code $\B\subseteq\Z_2^n$ is defined to be
\index{covering radius}
\beq
\rho(\B):=\max_{x\in\Z_2^n}\min_{b\in\B}\wt_H(b+x).
\eeq
\end{definition}
\vspace{2ex}
\par
In words, $\rho(\B)$ is the least nonnegative integer such that the spheres of radius $\rho(\B)$ around the codewords of $\B$ cover the space $\Z_2^n$. If $\B$ is linear, $\rho(\B)$ is equal to the maximum weight of the coset leaders of $\B$. Determining the covering radius of a binary code is generally nontrivial \cite{McLoughlin1984}. In spite of this, many results are known. For a good overview we refer to \cite{Cohen1997} and \cite{Brualdi1998}.
\par
Once and for all we fix the following parameters depending on $h$:
\beq
\lambda:=\begin{cases}
1 & \mbox{if}\quad h=1\\
2 & \mbox{if}\quad h>1
\end{cases}
\eeq
and
\beq
\epsilon:=\begin{cases}
1 & \mbox{if}\quad h=1\\
2^{2h-3}\sin^2\left(\frac{\pi}{2^h}\right) & \mbox{if}\quad h>1.
\end{cases}
\eeq
We are now in a position to state our main result, whose proof can be found in Section~\ref{sec:proofs}.
\begin{theorem}
\label{thm:LB_general}
Given a binary code $\B\subseteq\Z_2^n$ with covering radius \beq
\rho(\B)\le n\left(\frac{1}{\epsilon}-\frac{1}{2}\right),
\eeq
we have
\beq
\min_{\C\in E_h(\B)}\PMEPR(\C)\ge\frac{1}{\lambda n} \left[n(2-\epsilon)-2\,\epsilon\,\rho(\B)\right]^2.
\eeq
\end{theorem}
\vspace{2ex}
\par
The preceding lower bound is a decreasing function of $h$. However this decrease is bounded, and the corollary below states the asymptotic lower bound for $h\to\infty$.
\begin{corollary}
\label{cor:LB_arbitrary}
Let $\B\subseteq\Z_2^n$ be a binary code with covering radius 
\beq
\rho(\B)\le n\left(\frac{8}{\pi^2}-\frac{1}{2}\right).
\eeq
Then
\beq
\min_{\C\in E_\infty(\B)}\PMEPR(\C)\ge\frac{1}{2n}\left(n\,\frac{16-\pi^2}{8}-\frac{\pi^2}{4}\rho(\B)\right)^2.
\eeq
\end{corollary}
\vspace{2ex}
\begin{proof}
Recall that $\lambda=2$ for any $h>1$, and with the Taylor series 
\beqn
\label{eqn:Taylor_sin2}
\sin^2x=x^2+O(x^4)
\eeqn
we conclude
\beq
\lim_{h\to\infty}\epsilon=\lim_{h\to\infty}\;2^{2h-3}\sin^2\left(\frac{\pi}{2^h}\right)=\frac{\pi^2}{8}.
\eeq
The corollary is then immediate.
\end{proof}
\par
We note that the condition $\rho(\B)\le n/2$ (which is for $h>2$ slightly stronger than those in Theorem~\ref{thm:LB_general} and Corollary~\ref{cor:LB_arbitrary}) is always met if $\B$ has strength 1 \cite{Cohen1997}, \cite{Brualdi1998}, that is, each coordinate of $\B$ takes the values '$0$' and '$1$' equally often. In particular every linear code without a coordinate that is always zero has strength 1.
\par
We conclude from Theorem~\ref{thm:LB_general} that a necessary condition to obtain a code over $\Z_{2^h}$ with low PMEPR from a binary code is that the covering radius of the underlying binary code must be larger than a certain threshold. More specifically, to obtain a sequence of codes with PMEPR growth strictly less than order $n$ from a sequence of binary codes, the covering radius of the binary code must grow when $n$ increases. In particular, in order to achieve a constant PMEPR, the covering radius of the binary code $\B$ must satisfy
\beq
\rho(\B)\ge n\left(\frac{1}{\epsilon}-\frac{1}{2}\right)-O(\sqrt{n}),
\eeq
which can be replaced by the slightly stronger condition
\beqn
\label{eqn:ord_one}
\rho(\B)\ge \frac{n}{2}-O(\sqrt{n}).
\eeqn
We shall see in the next section that the preceding condition is satisfied for the duals of binary primitive BCH codes and for $r$th-order Reed--Muller codes. On the other hand, we will also see that Theorem~\ref{thm:LB_general} implies that the phase-shifted versions of many other codes have PMEPR growing linearly in $n$.


\section{Implications}
\label{sec:implications}

\subsection{Nonredundant BPSK Signaling}

Consider the code $\Z_2^n$. The associated PSK code is a BPSK code without redundancy. Since $\rho(\Z_2^n)=0$, Theorem~\ref{thm:LB_general} and Corollary~\ref{cor:LB_arbitrary} imply the following.
\begin{corollary}
\label{cor:BPSK}
For $h>1$ we have
\beq
\min_{\C\in E_h(\Z_2^n)}\PMEPR(\C)\ge \frac{(2-\epsilon)^2}{2}\cdot n
\eeq
and
\beq
\min_{\C\in E_\infty(\Z_2^n)}\PMEPR(\C)\ge 2n\left(1-\frac{\pi^2}{16}\right)^2\approx0.2936\cdot n.
\eeq
\end{corollary}
\vspace{2ex}
\par
In \cite{Wunder2002} Wunder and Boche proposed to use Newman phases \cite{Newman1965} to reduce the PMEPR for nonredundant BPSK signaling. Newman phases are given by
\beq
\varphi_i=\frac{\pi i^2}{n},\quad i=0,1,\dots,n-1.
\eeq
Applying the phase shift $\varphi_i$ to the $i$th coordinate of a BPSK code, we obtain a PSK code whose underlying code over $\Z_{2^\infty}$ is equivalent to $\Z_2^n$.
\par
\begin{figure}[t]
\centering
\includegraphics[width=0.65\columnwidth]{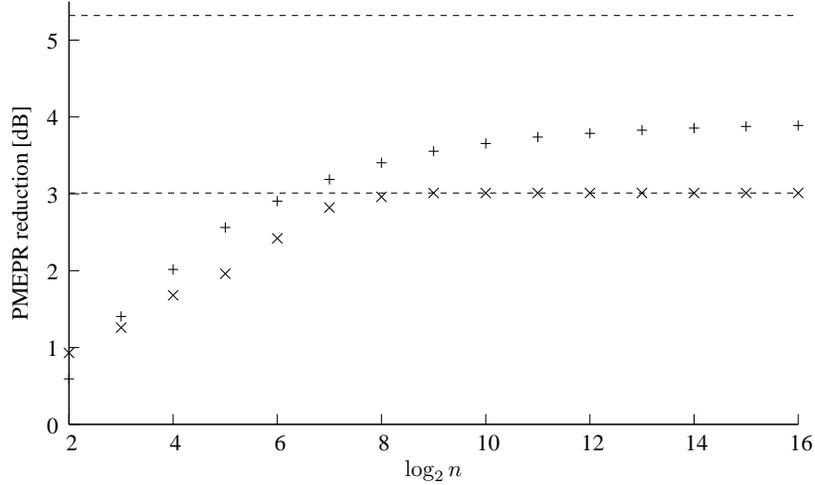}
\caption{PMEPR reduction for nonredundant BPSK signaling and $n$ carriers with Newman phases ($+$) and with Newman phases rounded to points in a QPSK constellation ($\times$), upper bounds (dashed lines)}
\label{fig:newman}
\end{figure}

For $n=16$ and $n=128$ in \cite{Wunder2002} the PMEPR could be reduced in this way from $12.04$~dB and $21.07$~dB to $9.97$~dB and $17.89$~dB, respectively. We have used the algorithm described in \cite{Wunder2002} to calculate the PMEPR reductions for lengths $n=2^m$, where $m$ ranges from $2$ to $16$. In addition we have rounded the Newman phases to the nearest points in $\{+1,+j,-1,-j\}$, i.e., the phase-shifted code is a QPSK code. The results are presented in Figure~\ref{fig:newman}. The figure also shows the corresponding bounds on the PMEPR reductions obtained with Corollary~\ref{cor:BPSK} ($-10\log_{10}0.5\approx 3.01$~dB for $h=2$ and $-10\log_{10} 0.2936\approx 5.32$~dB for $h\to\infty$).
By using rounded Newman phases the bound in Corollary~\ref{cor:BPSK} is attained for $m\ge 9$.  With the original Newman phases the PMEPR reduction seems to converge to approximately $3.9$~dB. Thus for large $n$ Newman phases approach our bound by about $1.4$~dB.

\subsection{Linear Codes}

Next we present a lower bound for the PMEPR of codes over $\Z_{2^h}$ that are equivalent to a given binary linear code as a function of the code rate.
\begin{corollary}
Let $\B\subseteq\Z_2^n$ be a binary linear code with rate 
\beq
R\ge \frac{3}{2}-\frac{1}{\epsilon}.
\eeq
Then
\beq
\min_{\C\in E_h(\B)}\PMEPR(\C)\ge\frac{4n}{\lambda}\left[1-\epsilon\left(\frac{3}{2}-R\right)\right]^2.
\eeq
\end{corollary}
\vspace{2ex}
\par
\begin{proof}
Let $k$ be the dimension of $\B$ such that $R=k/n$. The redundancy bound \cite{Cohen1997}, \cite{Brualdi1998} states 
\beqn
\label{eqn:redundancy_bound}
\rho(\B)\le n-k.
\eeqn
The condition $k/n\ge 3/2-1/\epsilon$ then implies that 
\beq
\rho(\B)\le n\left(\frac{1}{\epsilon}-\frac{1}{2}\right).
\eeq
Hence we can plug  (\ref{eqn:redundancy_bound}) into the inequality in Theorem~\ref{thm:LB_general}, which yields the desired expression.
\end{proof}
\par
The above corollary shows explicitly that it is impossible to construct a sequence of codes over $\Z_{2^h}$ from a sequence of binary linear codes with rate $R>3/2-1/\epsilon$ and PMEPR growth slower than linear in $n$.
\par
We remark that some improved bounds for linear codes can be obtained by taking bounds for the covering radius that are better than the redundancy bound, e.g., the Griesmer-like upper bound on the covering radius of linear codes \cite{Janwa1989}, which takes the minimum distance of the code into account.

\subsection{Binary Primitive BCH Codes and Their Duals}

In what follows we will analyze lower bounds on the PMEPR of codes that are equivalent to binary primitive $t$-error-correcting BCH codes, $\B(t,m)$, and their duals, $\B^\perp(t,m)$. These codes have length $2^m-1$. See, e.g., \cite{MacWilliams1977} for details on these codes.
\begin{corollary}
\label{cor:PMEPR_BCH}
Let $n=2^m-1$. Then we have
\beq
\min_{\C\in E_h(\B(t,m))}\PMEPR(\C)\ge\frac{1}{\lambda n}[n(2-\epsilon)-\epsilon(4t-2)]^2
\eeq
for $t=1$, $m\ge 2$ , for $t=2$, $m\ge 3$, for $t=3$, $m\ge 4$, and for arbitrary $t$ and $m\ge m_0$, where $m_0$ is finite and depends on $t$.
\end{corollary}
\begin{proof}
It is known that $\rho(\B(1,m))=1$ for $m\ge 2$ ($\B(1,m)$ is perfect), $\rho(\B(2,m))=3$ for $m\ge 3$ \cite{Gorenstein1960}, $\rho(\B(3,m))=5$ for $m\ge 4$ \cite{Helleseth1978}. More generally, it was proved in  \cite{Vladuts1989} that there exists an $m_0$ depending on $t$ such that $\rho(\B(t,m))=2t-1$ for all $m\ge m_0$. Putting these values into Theorem~\ref{thm:LB_general} yields the statement in the corollary.
\end{proof}
\begin{remark}
It has been shown in \cite{Cohen1997a} that $m_0$ satisfies
\beq
m_0\le2\log_2\left[(2t-1)!\;(2t-3)\right].
\eeq
\end{remark}
\par
The preceding corollary shows that every code in $E_h(\B(t,m))$ has PMEPR growing linearly in $n$. 
\par
Jones and Wilkinson \cite{Jones1996} applied a learning algorithm to obtain phase shifts for $\B(1,3)$ and $\B(1,4)$, the first nontrivial Hamming codes. As a result, \cite{Jones1996} reports binary phases that reduce the PMEPR from $8.45$~dB and $11.76$~dB to $5.53$~dB and $10.52$~dB, respectively. These values meet the lower bound in Corollary~\ref{cor:PMEPR_BCH}, and we conclude that the binary phase shifts computed in \cite{Jones1996} are the best possible. In \cite{Jones1996} the learning algorithm was also applied to find offsets over $\Z_8$. The respective reduced PMEPRs are $3.42$~dB and $8.47$~dB. For $\B(1,3)$ Corollary~\ref{cor:PMEPR_BCH} fails to provide a nontrivial lower bound. For $\B(1,4)$ Corollary~\ref{cor:PMEPR_BCH} yields the lower bound $5.30$~dB.
\par
Let us turn our attention to $\B^\perp(t,m)$.
\index{code!dual BCH}
\begin{corollary}
\label{cor:dual_BCH}
Let $n=2^m-1$. Then
\beq
\min_{\C\in E_1(\B^\perp(t,m))}\PMEPR(\C)\ge\frac{1}{n}\left[1+2(\sqrt{t}-\sqrt[6]{t})\sqrt{n-t-1}\right]^2.
\eeq
\end{corollary}
\vspace{2ex}
\par
\begin{proof}
The corollary follows from Theorem~\ref{thm:LB_general} and a result from \cite{Tietavainen1990}
\beq
\rho(\B^\perp(t,m))\le \frac{n-1}{2}-(\sqrt{t}-\sqrt[6]{t})\sqrt{n-t-1}.
\eeq
\end{proof}
\par
Observe that the covering radius of $\B^\perp(t,m)$ satisfies (\ref{eqn:ord_one}), and the lower bound on the PMEPR is asymptotically independent of the code length. However notice that the rate is strictly decreasing with increasing length. It is noteworthy that Paterson and Tarokh \cite{Paterson2000} obtained a bound on the PMEPR of the nonzero codewords of $\B^\perp(t,m)$, which has order $(\log n)^2$, and they speculated that by taking cosets of $\B^\perp(t,m)$ the PMEPR may be significantly further reduced. This conjecture is supported by Corollary~\ref{cor:dual_BCH}.

\subsection{Reed--Muller Codes}

Next we establish lower bounds on the PMEPR of codes that are equivalent to $r$th-order Reed--Muller codes of length $2^m$, $\RM(r,m)$. See, e.g., \cite{MacWilliams1977} for details on these codes. For $r=1$ we have the following.
\begin{corollary}
\label{cor:PMEPR_LB_RM}
The PMEPR of any code in $E_1(\RM(1,m))$ is at least 2 for odd $m\le 7$ and is lower bounded by 1 in all other cases. 
\end{corollary}
\begin{proof}
It was proved in \cite{Helleseth1978a} that
\beqn
\label{eqn:covering_radius_RM1}
2^{m-1}-2^{(m-1)/2}\le\rho(\RM(1,m))\le2^{m-1}-2^{m/2-1},
\eeqn
where the upper bound is tight when $m$ is even. The lower bound of 1 for the PMEPR follows then from Theorem~\ref{thm:LB_general} (for odd $m$ the lower bound can be tightened slightly by taking into account that $\rho(\RM(1,m))$ must be an integer). In the special case where $m$ is odd and $m\le 7$ it is known that the lower bound in (\ref{eqn:covering_radius_RM1}) is tight \cite{Berlekamp1972}, \cite{Mykkeltveit1980}.  Therefore Theorem~\ref{thm:LB_general} states that the PMEPR is lower bounded by 2 in this case.
\end{proof}
\par
Indeed Davis and Jedwab \cite{Davis1999} explicitly constructed $m!/2$ binary offsets for $\RM(1,m)$ that reduce the PMEPR to values not exceeding $2$ for any $m$. Hence the bound in Corollary~\ref{cor:PMEPR_LB_RM} is attained for odd $m\le 7$.
\par
Corollary~\ref{cor:PMEPR_LB_RM} only considers the case when $h=1$. However, even if $\rho(\RM(1,m))$ attains the lower bound in (\ref{eqn:covering_radius_RM1}), Theorem~\ref{thm:LB_general} yields the trivial lower bound 1 for the PMEPR of quaternary codes equivalent to $\RM(1,m)$ for all choices of $m$. 
\par
Let us turn our attention to Reed--Muller codes of arbitrary order. For $r>1$ general explicit results for the covering radius of $\RM(r,m)$ are unknown. For small $r$ and $m$ a few values and bounds are given in \cite[page 802]{Brualdi1998}. However good asymptotic bounds are known, which enable us to analyze the asymptotic behavior of Theorem~\ref{thm:LB_general} for all Reed--Muller codes of fixed order. Previous asymptotic bounds for $\rho(\RM(r,m))$ ($r>1$) have been recently improved in \cite{Carlet2007} to
\beq
\rho(\RM(r,m))\le 2^{m-1}-\frac{\sqrt{15}}{2}\left(\sqrt{2}+1\right)^{r-2}\cdot2^{m/2}+O(1).
\eeq
Observe that the covering radius satisfies (\ref{eqn:ord_one}). Hence for fixed $r$ Theorem~\ref{thm:LB_general} yields a lower bound that is asymptotically a constant. More specifically, we obtain the following.
\begin{corollary}
When $m$ tends to infinity, we have
\beq
\min_{\C\in E_1(\RM(r,m))}\PMEPR(\C)\ge 15\left(\sqrt{2}+1\right)^{2r-4}
\eeq
for $r>1$.
\end{corollary}
\par
This bound is indeed independent of the length of the code. But notice that for fixed $r$ the code rate tends to zero as the length of the code increases.

\subsection{Convolutional Codes}

Consider a convolutional encoder with $k_0$ binary inputs, $n_0$ binary outputs, and constraint length $\nu+1$. Suppose that the initial state of the encoder is arbitrary. The corresponding convolutional code consists of all possible output sequences of infinite length. The set of all binary output sequences of length $\ell n_0$ may be viewed as a linear block code $\C_\ell$ with generator matrix
\par
\beq
\begin{pmatrix}
g_\nu & 0 & \cdots & 0 & 0 & 0 & \cdots & 0\\
g_{\nu-1}  & g_\nu & \cdots & 0 & 0 & 0 & \cdots & 0\\
\multicolumn{8}{c}{\dotfill}\\
g_0  & g_1 & \cdots & g_\nu & 0 & 0 & \cdots & 0\\
0 & g_0  & \cdots & g_{\nu-1} & g_\nu & 0 & \cdots & 0\\
\multicolumn{8}{c}{\dotfill}\\
0 & 0 & \cdots & 0 & 0 & 0 & \cdots & g_0\\
\end{pmatrix},
\eeq
where $g_0, g_1,\dots,g_\nu$ are matrices of size $k_0\times n_0$ with elements in $\Z_2$, and $0$ is the all-zero matrix of the same size. Then $\C_\ell$ has length $n=\ell n_0$ and dimension $k=(\ell+\nu)k_0$. Note that the first $\nu k_0$ information bits are fed into the encoder to produce the initial state. This situation corresponds to the case where a block of $\ell$ subsequent symbols of a continuous stream of encoder outputs is used as a binary codeword of length $\ell n_0$ to modulate one OFDM symbol. So the initial state of this block is the final state of the previous block.
\par
The \emph{normalized covering radius} of a convolutional code $\C$ is defined to be \cite{Bocharova1993}, \cite{Calderbank1995}
\beq
\tilde\rho(\C):=\lim_{\ell\to\infty}\frac{\rho(\C_\ell)}{\ell n_0}.
\eeq
Bounds for the normalized covering radius of some convolutional codes can be found in \cite{Bocharova1993} and \cite{Calderbank1995}. The covering radius of $\C_\ell$ satisfies \cite{Bocharova1993}
\beq
\rho(\C_\ell)\le n\,\tilde \rho(\C).
\eeq
With Theorem~\ref{thm:LB_general} the next corollary is now immediate.
\begin{corollary}
With the notation as above, suppose that 
\beq
\tilde\rho(\C)\le \frac{1}{\epsilon}-\frac{1}{2}.
\eeq
Then
\beq
\min_{\D\in E_h(\C_\ell)}\PMEPR(\D)\ge\frac{4n}{\lambda}\left[1-\epsilon\left(\tilde\rho(\C)+\frac{1}{2}\right)\right]^2.
\eeq
\end{corollary}
\vspace{2ex}
\par
The preceding corollary states that, whenever $\tilde\rho(\C)\le 1/\epsilon-1/2$, the PMEPR of any code in $E_h(\C_\ell)$ grows linearly in $n$. We remark that for a particular $n$ the result can be tightened slightly by taking into account that the covering radius of a block code of finite length must be an integer. 
\par
As an example let us consider the scenario of binary signaling in the HiPerLAN/2 OFDM system \cite{Hiperlan2001}, which uses 48 carriers for data transmission. The employed convolutional code $\C$ is completely described by
\beq
(g_0,g_1,\dots,g_6)=((11),(01),(11),(11),(00),(10),(11)),
\eeq
so $\nu=6$, $k_0=1$, and $n_0=2$. Tarokh and Jafarkhani \cite{Tarokh2000} have computed quaternary, octary, and  $16$-ary phase shifts for $\C_{24}$ of length $48$. The respective reduced PMEPR is equal to $12.7$~dB, $12.6$~dB, and $12.4$~dB. We used a simple search algorithm in combination with the techniques reported in \cite{Wunder2002} to find a binary offset that reduces the PMEPR to $14.6$~dB. Notice that the PMEPR of the original code $\C_{24}$ is equal to $16.8$~dB since the code is linear and, therefore, contains the all-zero word.
\par
From \cite{Bocharova1993} we know that $\tilde \rho(\C)\le 1/5$, therefore, $\rho(\C_{24})\le 9$. However direct computation yields $\rho(\C_{24})=6$. Applying Theorem~\ref{thm:LB_general} directly, the PMEPR of any code in $E_h(\C_{24})$ is at least $14.3$~dB, $11.3$~dB, $8.4$~dB, and $7.4$~dB for $h=1,2,3,4$, respectively. We conclude that for $h=1$ a significant PMEPR reduction is precluded by our results and the reduced PMEPR approaches our bound by $0.3$~dB, for $h=2$ the PMEPR reduction obtained in \cite{Tarokh2000} is at least close to the best possible, and for $h=3,4$ the gap between the reduced PMEPR and our lower bound is about $4$--$5$~dB. This gap may have the following reasons: (i) Theorem~\ref{thm:LB_general} is not tight in this case, (ii) the optimization algorithm in \cite{Tarokh2000} converged to a local minimum, or (iii) a lower PMEPR can be obtained by applying another permutation to the coordinates of the code.


\section{Proofs}
\label{sec:proofs}

\subsection{Proof of Theorem~\ref{thm:LB_general} When $h=1$}

Here $\lambda=1$ and $\epsilon=1$. Let $a=(a_0,\dots,a_{n-1})\in\Z_2^n$. From (\ref{eqn:PMEPR_codeword}) and (\ref{eqn:OFDM_signal}) we have
\begin{align*}
\PMEPR(a)&\ge \frac{1}{n}|S_a(0)|^2\\
&=\frac{1}{n}\left|\sum_{i=0}^{n-1}(-1)^{a_i}\right|^2\\
&=\frac{1}{n}[n-2\wt_H(a)]^2.
\end{align*}
It follows that
\begin{align*}
\min_{\C\in E_1(\B)}\PMEPR(\C)&=\min_{\sigma}\min_{w\in\Z_2^n}\max_{b\in\B}\,\PMEPR(\sigma(b)+w)\\
&\ge\frac{1}{n}\min_{\sigma}\min_{w\in\Z_2^n}\max_{b\in\B}\,[n-2\wt_H(\sigma(b)+w)]^2\\
&=\frac{1}{n}\min_{w\in\Z_2^n}\max_{b\in\B}\,[n-2\wt_H(b+w)]^2.
\end{align*}
The condition $\rho(\B)\le n/2$ implies that $n-2\rho(\B)$ is nonnegative. Therefore we arrive at
\begin{align*}
\min_{\C\in E_1(\B)}\PMEPR(\C)&\ge\frac{1}{n}[n-2\max_{w\in\Z_2^n}\min_{b\in\B}\,\wt_H(b+w)]^2\\
&=\frac{1}{n}[n-2\rho(\B)]^2,
\end{align*}
and the theorem follows for $h=1$.\hfill\QED

\subsection{Proof of Theorem~\ref{thm:LB_general} When $h>1$}

We need some preliminaries in order to set out the proof. The \emph{Lee weight} of the word $a=(a_0,\dots,a_{n-1})\in\Z_{2^h}^n$ is defined to be
\beq
\wt_L(a):=\sum_{i=0}^{n-1}\min\left\{a_i,2^h-a_i\right\}.
\eeq
\begin{lemma}
\label{lem:Euclidean_Lee_dist}
For each $a,b\in\Z_{2^h}^n$ and $h>1$ we have
\beq
||\hat a-\hat b||^2\le 2^h\,\sin^2\left(\frac{\pi}{2^h}\right)\,\wt_L(a-b).
\eeq
\end{lemma}
\vspace{2ex}
\par

\begin{proof}
It suffices to prove the lemma for $n=1$. It is straightforward to establish that
\begin{align*}
||\hat a-\hat b||^2&=4\sin^2\left((a-b)\frac{\pi}{2^h}\right)\\
&=4\sin^2\left(\wt_L(a-b)\frac{\pi}{2^h}\right).
\end{align*}
Hence we have to show that
\beqn
\label{eqn:inequality_w}
\sin^2\left(w\,\frac{\pi}{2^h}\right)\le \frac{2^h}{4}\sin^2\left(\frac{\pi}{2^h}\right)\,w
\eeqn
holds for $h>1$ and $w=0,1,\dots,2^{h-1}$. When $h=2$,  equality in (\ref{eqn:inequality_w}) is easily proved by hand. For $h\ge 3$ the idea of the proof is as follows. Consider the function $f:\Rn\to\Rn$ with $f(x)=\sin^2(x\,\pi/2^h)$, which is a continuous equivalent of the LHS of (\ref{eqn:inequality_w}). First a tangent to the curve $f(x)$ at some $0<x_0\le2^{h-1}$ is constructed that passes through the origin. Then it is shown that for $0\le x \le 2^{h-1}$ and $h\ge 3$ the curve $f(x)$ is  upper bounded by this tangent and the slope of this tangent is at most
\beqn
\label{eqn:coeff_w}
\frac{2^h}{4}\sin^2\left(\frac{\pi}{2^h}\right),
\eeqn
which is the coefficient of $w$ in the RHS of (\ref{eqn:inequality_w}).
\par
The curve $f(x)$ has a local minimum at $x=0$ and a local maximum at $x=2^{h-1}$, and has exactly one inflection point between them. Since $x_0>0$, it is then obvious that the tangent must meet the curve $f(x)$ in the concave part at some $0<x_0\le 2^{h-1}$, where $x_0$ is uniquely determined. We conclude that $f(x)$ is upper bounded by the tangent for $0\le x\le 2^{h-1}$.
\par
Since the tangent passes through the origin, $x_0$ must satisfy
\beq
f(x_0)=f'(x_0)x_0,
\eeq
or equivalently,
\beqn
\label{eqn:solution_w_0}
\frac{2\pi x_0}{2^h}=\tan\left(\frac{\pi x_0}{2^h}\right).
\eeqn
Now we aim to show that $f'(x_0)$ is at most (\ref{eqn:coeff_w}), which is equivalent to showing that
\beqn
\label{eqn:inequality_final}
4\pi\sin\left(\frac{2\pi x_0}{2^h}\right)\le 2^{2h}\sin^2\left(\frac{\pi}{2^h}\right),
\eeqn
where $x_0$ satisfies (\ref{eqn:solution_w_0}). Notice that (\ref{eqn:solution_w_0}) is a transcendental equation, which generally does not have a closed-form solution. However we can apply numerical methods to solve (\ref{eqn:solution_w_0}), and find that the LHS of (\ref{eqn:inequality_final}) is upper bounded by $9.12$. The RHS of (\ref{eqn:inequality_final}) is an increasing function in $h$, and by using (\ref{eqn:Taylor_sin2}), we conclude that when $h\to\infty$, it converges to $\pi^2$, clearly greater than $9.12$. Hence there exists an integer $h_0$ so that (\ref{eqn:inequality_final}) is true for all $h\ge h_0$. Numerical analysis yields $h_0=3$.
\end{proof}
\par
The following definition was mentioned in \cite{Carlet1998} as a generalization of the classical Gray map.
\begin{definition}
\label{def:Gray_map}
For any integer $h>1$ we define the map $\phi:\Z_{2^h}\to \Z_2^{2^{h-1}}$ as follows. If $0\le a\le 2^{h-1}$, the image $\phi(a)$ is the binary word $(000\cdots 111)$ with Hamming weight equal to $a$. If $2^{h-1}<a<2^{h}$, the image $\phi(a)$ is the binary word $(111\cdots 000)$ with Hamming weight equal to $2^h-a$. We also define the maps $\beta,\gamma:\Z_{2^h}\to\Z_2^{2^{h-2}}$ such that for each $a\in\Z_{2^h}$
\beq
\phi(a)=(\beta(a),\gamma(a)).
\eeq
Finally the maps $\phi$, $\beta$, and $\gamma$ are extended in the obvious way to act on words in $\Z_{2^h}^n$.
\end{definition}
\par
In the special case when $h=2$, $\phi$ is the classical Gray map \cite{Carlet1998}. By virtue of its definition, $\phi$ is a weight-preserving map from $\Z_{2^h}^n$ supported by the Lee weight to $\Z_2^{2^{h-1}n}$ supported by the Hamming weight, i.e., for each $a\in\Z_{2^h}^n$ we have
\begin{align*}
\wt_L(a)&=\wt_H(\phi(a))\\
&=\wt_H(\beta(a))+\wt_H(\gamma(a)).
\end{align*}
We will need the following identities.
\begin{lemma}
\label{lem:Gray_map}
For $a\in\Z_{2^h}^n$ and $b\in\Z_2^n$ we have
\begin{align*}
(i)&\quad\beta(a-2^{h-2})=\overline{\gamma(a)},\quad\gamma(a-2^{h-2})=\beta(a)\\
(ii)&\quad\beta(a+2^{h-1}b)=\beta(a)+\beta(2^{h-1}b),
\end{align*}
where the addition of scalars is understood to be componentwise, and $\overline{\gamma(a)}$ denotes the complement of $\gamma(a)$.
\end{lemma}
\begin{proof}
The statements in the lemma are immediate consequences of the apparent fact that for $a,\delta\in\Z_{2^h}$ the binary word $\phi(a+\delta)$ is equal to a negacyclic shift by $\delta$ positions to the left of the binary word $\phi(a)$.
\end{proof}
\par
We are now in a position to complete the proof of Theorem~\ref{thm:LB_general}.
\par
\hspace{1em}{\itshape Proof of Theorem~\ref{thm:LB_general} When $h>1$:}
Now we have $\lambda=2$ and $\epsilon=2^{2h-3}\sin^2(\pi/2^h)$. Let $a=(a_0,\dots,a_{n-1})\in\Z_{2^h}^n$. It follows from (\ref{eqn:PMEPR_codeword}) and (\ref{eqn:OFDM_signal}) that
\begin{align*}
\PMEPR(a)&\ge\frac{1}{n}\left|S_a(0)\right|^2\\
&=\frac{1}{n}\left|\sum_{i=0}^{n-1}\hat a_i\right|^2.
\end{align*}
Using the easily verified identities
\begin{align*}
||\hat a-1||^2&=2n-2\Re\left\{\sum_{i=0}^{n-1}\hat a_i\right\}\\
||\hat a-j||^2&=2n-2\Im\left\{\sum_{i=0}^{n-1}\hat a_i\right\},
\end{align*}
where we interpret the addition of scalars to sequences componentwise, we obtain
\beq
\PMEPR(a)\ge\frac{1}{n}\left[\left(n-\frac{1}{2}||\hat a-1||^2\right)^2+\left(n-\frac{1}{2}||\hat a-j||^2\right)^2\right].
\eeq
The inequality
\begin{align*}
x^2+y^2&=\frac{1}{2}\left[(x+y)^2+(x-y)^2\right]\\
&\ge \frac{1}{2}(x+y)^2,\quad x,y\in\Rn
\end{align*}
yields
\beq
\PMEPR( a)\ge\frac{1}{2n}\left[2n-\frac{1}{2}\left(||\hat a-1||^2+||\hat a-j||^2\right)\right]^2.
\eeq
Applying Lemma~\ref{lem:Euclidean_Lee_dist} we arrive at
\beqn
\label{eqn:min_PMEPR_Lee}
\PMEPR(a)\ge\frac{1}{2n}\left[2n-\frac{\epsilon}{2^{h-2}}\left(\wt_L(a)+\wt_L(a-2^{h-2})\right)\right]^2,
\eeqn
provided that the expression inside the square is nonnegative. This will be assumed in the following and justified at the end of the proof. We now use the weight-preserving property of the generalized Gray map $\phi$ (cf. Definition~\ref{def:Gray_map}) and Lemma~\ref{lem:Gray_map}~(i) to establish
\begin{align*}
&\wt_L(a)+\wt_L(a-2^{h-2})\\
&\quad=\wt_H(\phi(a))+\wt_H(\phi(a-2^{h-2}))\\
&\quad=\wt_H(\beta(a))+\wt_H(\gamma(a))\\
&\qquad+\wt_H(\beta(a-2^{h-2}))+\wt_H(\gamma(a-2^{h-2}))\\
&\quad=2\wt_H(\beta(a))+\wt_H(\gamma(a))+\wt_H(\overline{\gamma(a)})\\
&\quad=2\wt_H(\beta(a))+2^{h-2}\,n,
\end{align*}
where $\overline{\gamma(a)}$ is the complement of $\gamma(a)$. Hence
\begin{align*}
\PMEPR(a)&\ge\frac{1}{2n}\left[2n-\frac{\epsilon}{2^{h-2}}(2\wt_H(\beta(a))+2^{h-2}n)\right]^2\\
&=\frac{1}{2n}\left[n(2-\epsilon)-\frac{\epsilon}{2^{h-3}}\wt_H(\beta(a))\right]^2.
\end{align*}
Indeed
\begin{align*}
&\min_{\C\in E_h(\B)}\PMEPR(\C)\\
&=\min_{\sigma}\min_{w\in\Z_{2^h}^n}\max_{b\in\B}\,\PMEPR(2^{h-1}\sigma(b)+w)\nonumber\\
&\ge\frac{1}{2n}\min_{\sigma}\min_{w\in\Z_{2^h}^n}\max_{b\in\B}\,
\left[n(2-\epsilon)-\frac{\epsilon}{2^{h-3}}\,\wt_H(\beta(w+2^{h-1}\sigma(b)))\right]^2\nonumber\\
&=\frac{1}{2n}\min_{w\in\Z_{2^h}^n}\max_{b\in\B}\,\left[n(2-\epsilon)-\frac{\epsilon}{2^{h-3}}\,\wt_H(\beta(w+2^{h-1}b))\right]^2\nonumber\\
&=\frac{1}{2n}\min_{w\in\Z_{2^h}^n}\max_{b\in\B}\,\left[n(2-\epsilon)-\frac{\epsilon}{2^{h-3}}\,\wt_H(\beta(w)+b')\right]^2,
\end{align*}
where Lemma~\ref{lem:Gray_map}~(ii) has been employed in the last step and $b'$ is given by
\beq
b'=\beta(2^{h-1}b)=(\underbrace{b\,b\,\cdots\,b}_{2^{h-2}\;\mbox{\scriptsize times}}).
\eeq
Recall that we imposed 
\beq
\rho(\B)\le n\left(\frac{1}{\epsilon}-\frac{1}{2}\right),
\eeq
which implies that the expression $n(2-\epsilon)-2\,\epsilon\,\rho(\B)$ is nonnegative. Hence we can write
\begin{align*}
&\min_{\C\in E_h(\B)}\PMEPR(\C)\\
&\ge\frac{1}{2n}\min_{u\in\Z_{2}^{2^{h-2}n}}\max_{b\in\B}\,\left[n(2-\epsilon)-\frac{\epsilon}{2^{h-3}}\wt_H(u+b')\right]^2\\
&=\frac{1}{2n}\min_{u\in\Z_{2}^n}\max_{b\in\B}\,\left[n(2-\epsilon)-2\,\epsilon\,\wt_H(u+b)\right]^2\\
&=\frac{1}{2n}\left[n(2-\epsilon)-2\,\epsilon\,\max_{u\in\Z_{2}^n}\min_{b\in\B}\, \wt_H(u+b)\right]^2\\
&=\frac{1}{2n}\left[n(2-\epsilon)-2\,\epsilon\,\rho(\B)\right]^2.
\end{align*}
Since the terms inside the squares in the preceding expressions are nonnegative, our assumption leading to (\ref{eqn:min_PMEPR_Lee}) holds, and the proof is completed.
\hfill\QED


\section{Conclusions and Open Problems}
\label{sec:conclusions}

In this paper we have analyzed the reduction of the PMEPR of a code when a permutation and a fixed phase shift is applied to its coordinates. A lower bound on the PMEPR for the case where the phase shifts are quantized (i.e., they are in the set $\{2\pi i/2^h\,|\,i=0,1,\dots,2^h-1\}$) was proved, and the asymptotic behavior for $h\to\infty$ was examined. The bound asserts that the achievable region of the PMEPR shrinks as the covering radius of the original code decreases.
\par
For $h=1$ and $h=2$ we exhibited examples where the lower bound in Theorem~\ref{thm:LB_general} is attained. Theorem~\ref{thm:LB_general} was also employed to show that most phase-shift designs from the literature are (nearly) optimal. It was demonstrated as well that for several code families of practical importance, including BCH codes and convolutional codes, a significant PMEPR reduction is ruled out by Theorem~\ref{thm:LB_general}. 
\par
We close with a discussion of some open problems and possible further research directions suggested by our work.
\par
We have identified codes for which the lower bound on the PMEPR is asymptotically independent of the length of the code. Namely these are the duals of the binary primitive BCH codes and Reed--Muller codes. Indeed Davis and Jedwab \cite{Davis1999} constructed $m!/2$ cosets of the first-order Reed--Muller code, $\RM(1,m)$, with PMEPR at most $2$. We ask: do there exist cosets of arbitrary Reed--Muller codes whose PMEPR is upper bounded by a constant? If so, find a way to construct them.
\par
We have analyzed lower bounds on the expression in (\ref{eqn:min_PMEPR}). Although such a bound can be used to rule out significant PMEPR reductions in certain cases, the bound does not claim the existence of $2^h$-ary codes equivalent to binary codes whose PMEPR is equal to or close to this bound. Therefore it would be interesting to find a good upper bound for (\ref{eqn:min_PMEPR}). Such a bound could provide results on the existence of good phase shifts.
\par
Finally we wish to restate the most essential (and most difficult) open problem within this context. Given $h$ and a binary code $\B\subseteq\Z_2^n$, find a code in $E_h(\B)$ whose PMEPR is equal to the value in (\ref{eqn:min_PMEPR}).

\vspace{-2ex}

\bibliographystyle{IEEEtranS}
\bibliography{IEEEabrv,references}

\end{document}